\documentclass[aps,prl,reprint,showpacs,showkeys,superscriptaddress]{revtex4-1}
\usepackage{amsmath,amssymb,bm,mathrsfs}
\usepackage{srcltx}
\usepackage{hyperref}

\begin{document}
\title{Unified Description of Nambu--Goldstone Bosons without Lorentz Invariance}

\author{Haruki Watanabe}
\email{hwatanabe@berkeley.edu}
\affiliation{Department of Physics, University of California,
  Berkeley, California 94720, USA}
\affiliation{Department of Physics, University of Tokyo, Hongo, Tokyo
  113-0033, Japan}

\author{Hitoshi Murayama}
\email{hitoshi@berkeley.edu, hitoshi.murayama@ipmu.jp}
\affiliation{Department of Physics, University of California,
  Berkeley, California 94720, USA} 
\affiliation{Theoretical Physics Group, Lawrence Berkeley National
  Laboratory, Berkeley, California 94720, USA} 
\affiliation{Kavli Institute for the Physics and Mathematics of the
  Universe (WPI), Todai Institutes for Advanced Study, University of Tokyo,
  Kashiwa 277-8583, Japan} 

\begin{abstract}
  Using the effective Lagrangian approach, we clarify general issues
  about Nambu--Goldstone bosons without Lorentz invariance.  We show
  how to count their number and study their dispersion relations.
  Their number is less than the number of broken generators when some
  of them form canonically conjugate pairs.  The pairing occurs when
  the generators have a nonzero expectation value of their commutator.
  For non-semi-simple algebras, central extensions are possible.
  The underlying geometry of the coset space in general is partially symplectic.
\end{abstract}

\maketitle

\paragraph{Introduction.}
---Spontaneous symmetry breaking (SSB) is ubiquitous in nature.  The
examples include magnets, superfluids, phonons, Bose--Einstein
condensates (BECs), neutron stars, and cosmological phase transitions.  When continuous and global
symmetries are spontaneously broken, the Nambu--Goldstone
theorem~\cite{Nambu:1961,Goldstone:1961,Goldstone:1962} ensures the
existence of gapless excitation modes, i.e., Nambu--Goldstone
bosons (NGBs).  Since the long-distance behavior of systems with SSB is dominated by NGBs, it is
clearly important to have general theorems on their number of degrees
of freedom and dispersion relations.

In Lorentz-invariant systems, the number of NGBs $n_{\mathrm{NGB}}$ is
always equal to the number of broken generators
$n_{\mathrm{BG}}$. All of them have the identical linear dispersion $\omega=c|k|$.
However, once we discard the Lorentz invariance, the situation varies from one system to another.

Until recently, systematic
studies on NGBs without Lorentz invariance have been limited. (See
Ref.~\cite{Brauner:2010} for a recent review.)
Nielsen and Chadha~\cite{Nielsen:1975} classified NGBs into two types:
type I (II) NGBs have dispersion relations proportional to odd (even)
powers of their momenta in the long-wavelength limit. They proved
$n_\mathrm{I}+2n_\mathrm{II}\geq n_\mathrm{BG}$,
where $n_\mathrm{I}$ ($n_\mathrm{II}$) is the number of type I (II) NGBs.
Sch\"afer~\textit{et al.}~\cite{Schafer:2001} showed that 
$n_\mathrm{NGB}$ is exactly equal to $n_\mathrm{BG}$
if $\langle0|[Q_i,Q_j]|0\rangle$ vanishes for all pairs of the symmetry generators $Q_i$.
A similar observation is given in
Ref.~\cite{Nambu:2004}.
Given these results, Brauner and one of us (H.~W.)~\cite{Watanabe:2011} conjectured
\begin{gather}
  n_\mathrm{BG}-n_\mathrm{NGB}=\frac12\mathrm{rank}\,\rho, \label{eq:conjecture}\\
  \rho_{ij}\equiv\lim_{\Omega\to\infty}\frac{-i}{\Omega}\langle0|[Q_i,Q_j]|0\rangle,
\label{eq:rho}
\end{gather}
where $\Omega$ is the spatial volume of the system.

In this Letter, we clarify these long-standing questions about the
NGBs in Lorentz-non-invariant systems by proving the conjecture and showing
the equality in the Nielsen-Chadha theorem with an improved definition
using effective Lagrangians $\mathcal{L}_{\mathrm{eff}}$. We also clarify how the central
extension of the Lie algebra makes a contribution to $\rho$~\footnote{For more details and extensive discussions, see H. Watanabe and H. Murayama (to be published).}.

\paragraph{Coset space.}
---When a symmetry group $G$ is spontaneously broken into its subgroup
$H$, the space of ground states form the coset space $G/H$ where two
elements of $G$ are identified if $g_1 = g_2 h$ for $^{\exists} h \in H$.  Every
point on this space is equivalent under the action of $G$, and we pick
one as the origin.  The unbroken group $H$ leaves the origin fixed,
while the broken symmetries move the origin to any other point.  The
infinitesimal action of $G$ is given in terms of vector fields ${\bold
  h}_i = h_i^{\phantom{i}a}\partial_a$ $(i =
1,\ldots,\mathrm{dim}\,G)$ on $G/H$, where $\partial_a=\frac{\partial}{\partial \pi^a}$ with the
local coordinate system $\{\pi^a\}$ $(a =
1,\ldots,n_{\mathrm{BG}}=\mathrm{dim}\,G-\mathrm{dim}\,H)$ around the origin.  
The infinitesimal transformations ${\bold h}_i$ satisfy the Lie
algebra $[{\bold h}_i, {\bold h}_j] = f^k_{\phantom{k}ij} {\bold
  h}_k$.  
We can always pick the
coordinate system such that $\pi^a$'s transform linearly under $H$,
namely, that ${\bold h}_i =\pi^bR^p(T_i)_b^{\phantom{b}a}\partial_a$,
where $R^p(T_i)$ is a representation of $H$~\cite{Coleman:1969}.  On the
other hand, the broken generators are realized nonlinearly, ${\bold h}_b=
h_b^{\phantom{b}a} (\pi)\partial_a$ with
$h_b^{\phantom{b}a} (0)\equiv X_b^{\phantom{b}a}$.  Since broken
generators form a basis of the tangent space at the origin, the matrix
$X$ must be full-rank and hence invertible.

The long-distance excitations are described by the NGB fields
$\pi^a(x)$ that map the space-time into $G/H$.  We now write down
its $\mathcal{L}_{\mathrm{eff}}$ in a systematic expansion in powers of
derivatives, because higher derivative terms are less important at
long distances.

\paragraph{Effective Lagrangians without Lorentz invariance.}
---We discuss the $\mathcal{L}_{\mathrm{eff}}$ for the NGB degrees of freedom following
Refs.~\cite{Leutwyler:1994nonrel, Leutwyler:1994rel}.  Under \textit{global}\/ symmetry $G$, the NGBs transform as
$\delta \pi^a = \theta^i h_i^{\phantom{i}a}$ where $\theta^i$ are
infinitesimal parameters.  However, we do not make $\theta^i$ local
(gauge) unlike in these papers because it puts unnecessary
restrictions on possible types of symmetries and their realizations,
as we will see below.

It is well known that a symmetry transformation can change the Lagrangian density by
a total derivative.  The examples include space-time translations, supersymmetry, and gauge symmetry in the Chern--Simons
theory~\cite{Murayama:1989we}.  We allow for this possibility in the $\mathcal{L}_{\mathrm{eff}}$ of
the NGB fields.  We assume spatial translational invariance and
rotational invariance at sufficiently long distances in the continuum
limit, while we can still discuss their SSB.  

If Lorentz invariant, the $\mathcal{L}_{\mathrm{eff}}$ is highly constrained,
\begin{eqnarray}
\mathcal{L}_\mathrm{eff} &=&
\frac{1}{2}g_{ab}(\pi)\partial_\mu\pi^a\partial^\mu\pi^b 
+O(\partial_\mu^4).
\end{eqnarray}
The invariance of the Lagrangian under $G$ requires that $g_{ab}$ is
a $G$-invariant metric on $G/H$, namely $\partial_cg_{ab}h_i^{\phantom{i}c} + g_{ac}\partial_bh_i^{\phantom{i}c}
+g_{cb}\partial_ah_i^{\phantom{i}c}=0$. 
When the coordinates $\pi^a$ are reducible under $H$, the
metric $g$ is a direct sum of irreducible
components $g_{ab}= \sum_p F_p^2 \delta^p_{ab}$ where
$\delta^p_{ab}$ vanishes outside the irreducible representation $p$
with arbitrary constants $F_p$ for each of them.

On the other hand, once we drop Lorentz invariance, 
the general $\mathcal{L}_{\mathrm{eff}}$ has substantially more freedom,
\begin{eqnarray}
\mathcal{L}_\mathrm{eff} &=& c_a(\pi)\dot{\pi}^a
+\frac{1}{2}\bar{g}_{ab}(\pi)\dot{\pi}^a\dot{\pi}^b
-\frac{1}{2}g_{ab}(\pi)\partial_r\pi^a\partial_r\pi^b \nonumber \\
& & +O(\partial_t^3,\partial_t\partial_r^2,\partial_r^4),
\label{eq:effective}
\end{eqnarray}
where $\bar{g}_{ab}$ is also $G$ invariant.  Here and hereafter,
$r=1,\ldots,d$ refers to spatial directions.

Note that the spatial isotropy does not allow terms with first
derivatives in space in the $\mathcal{L}_{\mathrm{eff}}$.  Therefore, the spatial derivatives
always start with at least the second power $O(\partial_r^2)$.
(Actually, it is not critical for us whether there are terms of
$O(\partial_r^2)$;  it may as well start at $O(\partial_r^4)$ without
affecting our results, as we will see below.)  

The Lagrangian density changes by a total derivative under the
infinitesimal transformation $\delta \pi^a = \theta^i
h_i^{\phantom{i}a}$ \textit{iff}\/
\begin{gather}
  \left(\partial_b c_a-\partial_a
  c_b\right) h_i^{\phantom{k}b} = \partial_ae_i.
  \label{eq:hw}
\end{gather}
The functions $e_i(\pi)$ introduced in this way are actually related to the charge densities of the system. By paying attention to the variation of the Lagrangian by the surface term 
\begin{equation}
\delta \mathcal{L}_{\mathrm{eff}} = \theta^i \partial_t (c_a h_i^{\phantom{i}a}+ e_i),
\label{eq:totalderivative}
\end{equation}
we can derive the Noether current for the global symmetry $j^0_i = e_i -
\bar{g}_{ab} h_i^{\phantom{i}a} \dot{\pi}^b$.  
Since the ground state is time independent $\dot{\pi}^b = 0$,
\begin{equation}
e_i(0)=\langle 0|j_i^0(x)|0\rangle.
\label{eq:ei}
\end{equation}
It must vanish in the Lorentz-invariant case, which
can be understood as the special situation where $c_a$ and $e_i$
vanish, and $g_{ab}=c^2 \bar{g}_{ab}$.

Before presenting the proof, we explain the advantage in not gauging the
symmetry. A tedious calculation verifies
$\partial_b(h_i^{\phantom{i}a}\partial_ae_j-f^k_{\phantom{k}ij} e_k)=0$, with a general
solution,
\begin{equation}
  h_i^{\phantom{i}a}\partial_a e_j=f^k_{\phantom{k}ij} e_k+ c_{ij}. 
\label{eq:he}
\end{equation}
Therefore, $e_i(\pi)$'s transform as the adjoint representation under
$G$, up to possible integration constants $c_{ij} = - c_{ji}$. These constants play important roles as seen below.

In the presence of such constants, the global symmetry cannot be gauged~\cite{Leutwyler:1994nonrel}.
This is reminiscent of the Wess--Zumino term that also changes
by a surface term under a global symmetry and produces an anomaly
upon gauging~\cite{Wess:1971,Witten:1983}.  It is known that the constants
can be chosen to vanish with suitable definitions of $e_i$ for
semisimple Lie algebras, while a nontrivial second cohomology
of the Lie algebra presents an obstruction~\cite{Woodhouse:1992de}.

\paragraph{Proof of the conjecture.}
---The basic point to show is that when $\rho_{ij} \neq 0$, the NGB
fields for the generators $i$ and $j$ are
\textit{canonically conjugate}\/ to each other.

From Eq.~\eqref{eq:ei} and the assumed translational
symmetry, the formula for $\rho$ in Eq.~\eqref{eq:rho} is reduced to
\begin{equation}
  \rho_{ij} =-i \langle 0 | [Q_i, j^0_j] | 0 \rangle
  = h_i^{\phantom{i}a}\partial_a e_j\big|_{\pi=0}.
\end{equation}
Obviously, this must vanish for unbroken generators by definition.
Combining this with Eq.~\eqref{eq:hw}, we have
\begin{equation}
h_i^{\phantom{i}a}h_j^{\phantom{i}b}
\left(\partial_bc_a-\partial_a c_b\right)\big|_{\pi=0}=\rho_{ij}.
\label{eq:de}
\end{equation}
We now solve this differential equation around the origin. The Taylor
expansion of $c_a(\pi)$ can be written as
$c_a(\pi)=c_a(0)+(S_{ab}+A_{ab})\pi^b+O(\pi^2)$, where $S_{ab}$ and
$A_{ab}$ stand for the symmetric and antisymmetric parts of the
derivative $\partial_b c_a|_{\pi=0}$. Obviously
$c_a(0)$ and $S_{ab}$ lead to only total derivative terms in the $\mathcal{L}_{\mathrm{eff}}$ and thus will be dropped later:
\begin{equation}
  c_a(\pi)\dot{\pi}^a
  =A_{ab}\dot{\pi}^a\pi^b
  +\partial_t
  \left[c_a(0)\pi^a+\frac{1}{2}S_{ab}\pi^a\pi^b\right]+O(\pi^3).
\end{equation}
The equation for the antisymmetric part
$2X_c^{\phantom{b}a}X_d^{\phantom{b}b}A_{ab}=\rho_{cd}$ has a unique
solution which gives
\begin{equation}
c_a(\pi)\dot{\pi}^a=\frac{1}{2}\rho_{ab}\dot{\tilde{\pi}}^a\tilde{\pi}^b+O(\tilde{\pi}^3),
\end{equation}
where $\tilde{\pi}^a\equiv\pi^b(X^{-1})_b^{\,\,a}$. Since the matrix $\rho$ is real and antisymmetric, we can always
transform it into the following form by a suitable orthogonal
transformation $\tilde{Q}_i=O_{ij}Q_j$:
\begin{equation}
\rho=
\begin{pmatrix}
M_1&&&&&\\
&\ddots&&&&\\
&&M_m&&&\\
&&&0&&\\
&&&&\ddots&\\
&&&&&0
\end{pmatrix}
,\,\,
M_\alpha=
\begin{pmatrix}
0&\lambda_\alpha\\
-\lambda_\alpha&0
\end{pmatrix}.
\end{equation}
Here, $\lambda_\alpha\neq0$ for
$\alpha=1,\ldots,m=\frac{1}{2}\mathrm{rank}\,\rho$, while the
remaining elements identically vanish.  

The most important step in the proof is to write down the explicit expression of the $\mathcal{L}_{\mathrm{eff}}$ in Eq.~\eqref{eq:effective},
\begin{equation}
  c_a(\pi)\dot{\pi}^a
  =\sum_{\alpha=1}^m\dfrac{1}{2}\lambda_{\alpha}
  (\tilde{\pi}^{2\alpha}\dot{\tilde{\pi}}^{2\alpha-1}
  -\dot{\tilde{\pi}}^{2\alpha}\tilde{\pi}^{2\alpha-1}),
  \label{eq:c}
\end{equation}
which is in the familiar form of the Lagrangian on the phase space
$L=p_i\dot{q}^i-H$~\footnote{Darboux's theorem says one can choose a local
coordinate system such that higher order terms $O(\tilde{\pi}^3)$ vanish.}.
Namely, $\tilde{\pi}^{2\alpha-1}$ and $\tilde{\pi}^{2\alpha}$ are canonically conjugate
variables, and they together represent one degree of freedom rather than two degrees of freedom.
Hereafter we call the first set of $\tilde{\pi}^a$'s $(a=1,
\ldots 2m)$ type B, and the rest type A. Hence,
$n_\mathrm{A}+2n_\mathrm{B}=n_\mathrm{BG}$ with
$n_\mathrm{A}=n_\mathrm{BG}-2m$ and $n_\mathrm{B}=m$. Thus we proved the conjecture Eq.~\eqref{eq:conjecture}.

The definition of a degree of freedom here 
is the conventional one in physics; i.e., one needs
to specify both the instantaneous value and its time derivative for
each degree of freedom as initial conditions.  This definition does
not depend on the terms with spatial derivatives in the Lagrangian.

Now we are in the position to prove that the equality is satisfied in the Nielsen--Chadha theorem if the term with two spatial derivatives
exists with a nondegenerate metric $g_{ab}$.  Then Eq.~\eqref{eq:effective} implies that the type A NGB fields have
linear dispersion relations $\omega\propto k$, while the type B
NGB fields have quadratic dispersions
$\omega\propto k^2$. In this case, our type A (B) coincides
with their type I (II), respectively, and the Nielsen--Chadha
inequality is saturated.

On the other hand, if we allow the second-order term $O(\partial_r^2)$ to
vanish accidentally but the fourth-order term $O(\partial_r^4)$ to exist, the
unpaired (type A) NGBs happen to have a quadratic dispersion ($\omega^2 \propto k^4$, and hence type II) yet count as
independent degrees of freedom each~\cite{Watanabe:2011}.  Therefore, the Nielsen--Chadha theorem is still an inequality in general. In contrast, our distinction between type A and type B NGBs is
clearly determined by the first two time derivatives, and defines the
number of degrees of freedom unambiguously.  
Therefore, the classification
between odd and even powers in the dispersion relation is not an
essential one, and our theorem is stronger than that by Nielsen and Chadha.

Note that the Lagrangian formalism is mandatory in our discussion,
because the presence of the first-order derivatives in time
essentially affects the definition of the canonical momentum, while a
Hamiltonian is written with a fixed definition of the canonical
momentum.

\paragraph{Examples.}
---The simplest and most famous example of a type B NGB is the
Heisenberg ferromagnet $H=-J\sum_{\langle i,j\rangle}
\bm{s}_i \cdot \bm{s}_j$ with $J>0$ on a $d$-dimensional square
lattice ($d>1$). In this case, the original symmetry group
$\mathrm{O}(3)$ is spontaneously broken down into the subgroup
$\mathrm{O}(2)$. The coset space is $\mathrm{O}(3)/\mathrm{O}(2)\cong
S^2$.  We assume that the ground state has all spins lined up along
the positive $z$ direction without a lack of generality.  Even though
there are two broken generators, there is only one NGB with the
quadratic dispersion relation $\omega \propto k^2$.

The coset space can be parametrized as $(n_x,n_y,n_z)=\big(\pi^1,\pi^2,\sqrt{1-(\pi^1)^2-(\pi^2)^2}\big)$. The $\mathrm{O}(3)$ transformation $h_{i}^{\phantom{i}a}=\epsilon_{iaj}n_j$ ($i,j=x,y,z$; $a=1,2$) is realized linearly for the unbroken generator $h_z^{\phantom{z}a}(\pi)=\epsilon_{ab}\pi^b$, while nonlinearly for broken ones $X_a^{\phantom{a}b}=\epsilon_{ab}$. One can show that the $\mathcal{L}_{\mathrm{eff}}$ consistent with the $\mathrm{O}(3)$ symmetry up to $O(\partial^2$) is
\begin{equation}
  \mathcal{L}_\mathrm{eff} = m\frac{
    n_y\dot{n}_x-n_x\dot{n}_y}{1+n_z} 
  +\frac{1}{2}\bar{F}^2 \dot{\bm{n}}^2
  -\frac{1}{2}F^2\partial_r\bm{n}\partial_r\bm{n} .
  \label{eq:ferro}
\end{equation}
Comparing to Eqs.~\eqref{eq:effective}~and~\eqref{eq:hw}, we can  read off $c_a$ and $e_i$ as
$c_1=\frac{mn_y}{1+n_z}$, $c_2=-\frac{mn_x}{1+n_z}$ and $e_i=m n_i$. Hence $m=\langle j^0_z\rangle$ represents the magnetization of the ground state.
It is clear that there is only one type B NGB because $\pi^1$ and $\pi^2$
are canonically conjugate to each other, with a quadratic dispersion
$\omega \propto k^2$.

However, for an antiferromagnet, $J<0$, the overall magnetization
cancels between sublattices, and therefore $e_i = 0$, which in
turn requires $c_a = 0$.  As a consequence, the lowest order term in
the time derivative expansion has two powers, and we find that
both $\pi^1$ and $\pi^2$ represent independent type A NGBs 
with linear dispersions $\omega
\propto |k|$.  The generalization to the ferrimagnetic case is straightforward.

Another example is the spontaneously broken translational invariance
that leads to acoustic phonons in an isotropic medium~\cite{Landau:1986}.
The displacement vector $\bm{u} (x)$
represents the NGBs under the spatial
translation $\bm{u} \rightarrow \bm{u} + \bm{\theta}$, hence $G={\mathbb R}^3$
and $H=0$. Then with $\mathrm{O}(3)$ symmetry of spatial
rotations, the most general form of the continuum $\mathcal{L}_{\mathrm{eff}}$ is
\begin{equation}
  \mathcal{L}_{\mathrm{eff}} = \frac{1}{2} \dot{\bm{u}}^2
  - \frac{c_\ell^2}{2} (\bm{\nabla} \cdot \bm{u})^2 
  - \frac{c_t^2}{2} (\bm{\nabla} \times \bm{u})^2.
\end{equation}
We recover the usual result of one longitudinal and two transverse
phonons with linear dispersions $\omega=c_\ell k$ and $\omega=c_t k$, respectively (type A).
When the $\mathrm{O}(3)$ symmetry is reduced to $\mathrm{SO}(2)\times
{\mathbb Z}_2$ for rotation in the $xy$ plane and the reflection $z
\rightarrow -z$, there are considerably more terms one can write down.
Using the notation $\psi = u_x + i u_y$,
$\partial=\frac{1}{2}(\partial_x - i \partial_y)$, and $\bar{\psi}$
and $\bar{\partial}$ for their complex conjugates, the most general
$\mathcal{L}_{\mathrm{eff}}$ is
\begin{eqnarray}
  \lefteqn{
    {\cal L}_{\rm eff}
    = \dfrac{ic_{xy}}{2} \bar{\psi} \dot{\psi}
    + \frac{1}{2} \dot{u}_z^2 + \dot{\bar{\psi}} \dot{\psi}
    - F_{0}^2 (\bar{\partial} \psi) (\partial \bar{\psi})
    - \frac{1}{2} F_{1}^2 (\partial_z u_z)^2 
  } \nonumber \\
  & &
  - (\bar{\partial} u_z, \partial_z \psi)
  \left(
    \begin{array}{cc}
      F_{2}^2 & F_{3}^2 \\ F_{3}^{2} & F_{4}^2
    \end{array}
  \right)
  \left(
    \begin{array}{c}
      \partial u_z \\ \partial_z \bar{\psi}
    \end{array}
  \right) 
  - \frac{1}{2} ( F_{5}^2 (\partial \psi)^2 + c.c.) . \nonumber \\
\end{eqnarray}
With $c_{xy}\neq 0$, we find there is one type A NGB with a linear
dispersion, and one type B NGB with a quadratic dispersion.  The first
term $\frac{ic_{xy}}{2} \bar{\psi} \dot{\psi} =\frac12 c_{xy} (u_y \dot{u}_x-u_x \dot{u}_y)$
implies
\begin{equation}
  \rho_{xy} =-i\langle 0 | [P_x, j_y^0] | 0\rangle
  = c_{xy} \neq 0.
\end{equation}
Namely, this Lie algebra is a \textit{central extension}\/ of the
Abelian algebra of the translation generators, i.e., $[P_i,
P_j] = c_{ij}\Omega$.  As pointed out in Ref.~\cite{Watanabe:2012}, when the medium is electrically charged, an
external magnetic field along the $z$ axis precisely leads to this behavior with $c_{xy}=2\omega_c$ (the cyclotron frequency), because the
gauge-invariant translations in a magnetic field are generated by $P_i = -i\hbar \partial_i - \frac{e}{c} A_i$, which satisfy $\langle 0|[P_x , P_y]|0\rangle = i\frac{\hbar e}{c} B_z
N$ with the number of particles $N$. This would not be possible
with the gauged $\mathcal{L}_{\mathrm{eff}}$ in Ref.~\cite{Leutwyler:1994nonrel} that does not allow for the central extension.

As a more nontrivial example, let us consider a spinor BEC with $F=1$.
The symmetry group is $G=\mathrm{SO}(3) \times \mathrm{U}(1)$, where
$\mathrm{SO}(3)$ rotates three components of $F=1$ states, while
$\mathrm{U}(1)$ symmetry gives the number conservation.  The Lagrangian is written using a
three-component complex Schr\"odinger field $\psi$,
\begin{equation}
  \mathcal{L}=i\hbar\psi^\dagger\dot{\psi}
  -\dfrac{\hbar^2}{2m}\partial_r\psi^\dagger\partial_r\psi
  +\mu\psi^\dagger\psi-\dfrac{\lambda}{4}(\psi^\dagger\psi)^2
  -\dfrac{\kappa}{4}|\psi^T \psi|^2.
  \label{eq:BEC}
\end{equation}
Since the potential reads $\frac{\lambda+\kappa}{4}\hat{n}^2-\mu \hat{n}-\frac{\kappa}{4}\hat{\bm{S}}^2$ ($\hat{n}\equiv \psi^{\dagger}\psi$, $\hat{\bm{S}}\equiv\psi^{\dagger}\bm{S}\psi$ and $\bm{S}$ is the 3 by 3 spin matrix), we identify two possibilities for condensates
\begin{equation}
  \psi =v_p(0,0,1)^T\quad
  \mbox{or} \quad \frac{v_f}{\sqrt{2}}(1,i,0)^T
\end{equation}
for ``polar'' $(-\lambda < \kappa < 0)$ or ``ferromagnetic''
$(\kappa>0)$ states, where $v_p=\sqrt{\frac{2\mu}{\lambda+\kappa}}$ and $v_f= \sqrt{\frac{2\mu}{\lambda}}$~\cite{Ho:1998zz}.  The magnetization density is
given by $e_i(0)=\langle j_i^0 \rangle=- i \hbar \epsilon_{ijk} \psi_j^*
\psi_k$.  In the polar case, there is no net magnetization, and the
symmetry is broken into $H=\mathrm{SO}(2) \subset \mathrm{SO}(3)$.  For
the ferromagnetic case, there is a net magnetization $e_z(0)=\hbar v_f^2$, and the symmetry is broken into the diagonal subgroup $H$
of $\mathrm{U}(1)$ and $\mathrm{SO}(2) \subset \mathrm{SO}(3)$.
Therefore, the unbroken symmetry is the same for both cases
[$H=\mathrm{SO}(2)=\mathrm{U}(1)$], yet we see three type A NGBs for
the polar case while one type A and one type B NGB for the
ferromagnetic case as shown below.

For the polar case, we parameterize $\psi$ as
\begin{equation}
  \psi = (v_p + h) e^{i\theta} (\vec{n} + i \vec{\chi}), \quad
  \vec{n}^2 = 1, \quad \vec{\chi} \perp \vec{n}.
\end{equation}
After integrating out the gapped modes $h$ and $\vec{\chi}$, the
Lagrangian (\ref{eq:BEC}) becomes
\begin{equation}
  \mathcal{L}_{\mathrm{eff}} = \frac{\hbar^2}{\lambda+\kappa} \dot{\theta}^2
  + \frac{\hbar^2}{|\kappa|} \dot{\vec{n}}^2 
  - \frac{\hbar^2 v_p^2}{2m} \left[ (\partial_r \theta)^2
    + (\partial_r \vec{n})^2
    \right].
\end{equation}
We do find three type A NGBs with linear dispersions.

For the ferromagnetic case, we parameterize $\psi$ as
\begin{equation}
  \psi = (v_f+h) 
  \dfrac{e^{i\theta}}{\sqrt{2}(1+z^* z)}\begin{pmatrix}1-z^2\\i(1+z^2)\\2z\end{pmatrix}.
\label{eq:RP3}
\end{equation}
After integrating out $h$, we find
\begin{eqnarray}
  \lefteqn{
  \mathcal{L}_{\mathrm{eff}} = \hbar v^2_f i 
  \dfrac{z^* \dot{z} - \dot{z}^* z}{1+z^* z}
  + \frac{\hbar^2}{\lambda} \left( \dot{\theta} -i \dfrac{z^* \dot{z}-\dot{z}^*
    z}{1+z^* z}\right)^2 } \nonumber \\
   & & -\frac{\hbar^2 v^2_f}{2m} \left[
    \left(\partial_r \theta -i \dfrac{z^* \partial_r{z}-\partial_r{z}^*
    z}{1+z^* z}\right)^2 
    + \dfrac{2\partial_r z^* \partial_r z}{(1+z^* z)^2}\right].\quad
\label{eq:LeffRP3}
\end{eqnarray}
Clearly, $z$ and $z^*$ are canonically conjugate to each other,
representing one type B NGB with a quadratic dispersion, while
$\theta$ represents one type A NGB with a linear dispersion.

\paragraph{Underlying geometry.}
---Having demonstrated our theorem
Eq.~\eqref{eq:conjecture} at work in very different
examples, we now study the underlying geometry.  Usually, canonically
conjugate pairs in mechanics (such as type B NGBs) imply a \textit{symplectic structure}\/ mathematically, which requires an
even-dimensional manifold $M$, and if closed, a nontrivial second de
Rham cohomology $H^2(M) \neq 0$.  However, we have seen in the last
two examples that type A and type B NGBs can coexist on an
odd-dimensional $M$ with $H^2(M)=0$.  This puzzle can be solved as follows.

The time integral of the first term in Eq.~\eqref{eq:effective}
defines a one-form $c= c_a d \pi^a$ on the coset space, and
its exterior derivative gives a closed two-form $\sigma=dc$.
Using the coordinates in Eq.~\eqref{eq:c}, $\sigma
=\sum_{\alpha=1}^m\lambda_\alpha
d\tilde{\pi}^{2\alpha}\wedge d\tilde{\pi}^{2\alpha-1}$
for $m$ type B NGBs, which resembles a symplectic two-form.  However,
type A NGB fields for the remaining $n_{\mathrm{BG}}-2m$ broken
generators do not have terms with first order in time derivatives, and
hence do not take part in $\sigma$.  Therefore, $\sigma$ has a
constant rank but is degenerate, and hence is not a symplectic structure in
the usual sense.

This kind of a partially symplectic (or \textit{presymplectic}\/~\cite{Woodhouse:1992de}) structure is possible on
a coset space by considering the following fiber bundle,
$F\hookrightarrow G/H\stackrel{\pi}{\rightarrow}B$, where the base
space $B = G/(H\times F)$ is symplectic. The fiber $F$ is a subgroup
of $G$ that commutes with $H$. The symplectic structure $\omega$ on
$B$ is pulled back to $G/H$ as $\sigma=\pi^* \omega$. Since
$d\omega=0$ on $B$ implies $d\sigma = 0$ on
$G/H$, we can always find a one-form $c$ such that $dc =
\sigma$ locally on $G/H$, which appears in $\mathcal{L}_{\mathrm{eff}}$. Type B NGBs
live on the symplectic base manifold $B$, whose coordinates form
canonically conjugate pairs, while the type A NGBs live on the fiber
$F$, each coordinate representing an independent NGB.  The
type A and type B NGBs can coexist on $G/H$ in this fashion.

The Heisenberg ferromagnetic model has the coset space $S^2 = {\mathbb
  C}{\mathrm P}^1$ which is K\"ahler and hence symplectic, with one
type B NGB.  On the other hand, the spinor BEC example in its
ferromagnetic state has $G/H={\mathbb R}{\mathrm P}^3$ which is not
symplectic.  The last term in Eq.~\eqref{eq:LeffRP3} is nothing but
the Fubini--Study metric on $S^2 = \mathbb{C}\mathrm{P}^{1}$ which is
K\"ahler and hence symplectic.  The first term in
Eq.~\eqref{eq:LeffRP3} defines the one-form $c$ whose exterior
derivative ${\mathrm d}c$ gives precisely the K\"ahler form associated
with the metric up to normalization.  However $\theta$ is an
orthogonal direction with no connection to the symplectic structure.
We can define the projection $\pi:{\mathbb R}{\mathrm P}^3 \rightarrow
S^2$ simply by eliminating the $\theta$ coordinate.  It shows the
structure of a fiber bundle $\mathrm{U}(1) \hookrightarrow {\mathbb R}{\mathrm
  P}^3 \stackrel{\pi}{\rightarrow} \mathbb{C}\mathrm{P}^{1}$, which is
the well-known Hopf fibration (the difference between $S^3$ and
${\mathbb R}{\mathrm P}^3 = S^3/{\mathbb Z}_2$ is not essential here).
The phonons in the magnetic field also show a partially symplectic
structure.

In fact, it is always possible to find such a symplectic manifold $B$
if $G$ is compact and semisimple, thanks to Borel's theorem~\cite{Borel:1954}.
Generalizations to non-semi-simple
groups would be an interesting future direction in mathematics.

\paragraph{Final remarks.}
---In this Letter, we exclusively focused on true NGBs. We do not
regard pseudo-NGBs~\cite{Weinberg:1972} as NGBs, since they do not correspond to the broken
symmetries and tend to
acquire mass corrections.  Also, we assumed that there are no gapless excitations other
than NGBs; especially, this assumption fails when there is a Fermi
surface on the ground state. Taking such degrees of freedom into
account would be another interesting future direction.

\acknowledgments We thank T.~Brauner, D.~Stamper-Kurn, M.~Ueda, and K.~Hori for
useful discussions.  H.~W.~is also grateful to T.~Hayata and H.~M.~to T.~Milanov.
The work of H.~M.~was supported in part by the U.S. DOE under Contract No.~DEAC03-76SF00098, by the NSF under Grant No.~PHY-1002399, by the JSPS Grant (C) No.~23540289, by the
FIRST program Subaru Measurements of Images and
Redshifts (SuMIRe), CSTP, and by WPI, MEXT, Japan.

\bibliography{references}
\end{document}